\documentclass[12pt]{article}

\usepackage{bm,amsmath,amssymb,latexsym,graphicx,euscript,multirow,color,url,verbatim}
\usepackage{epsf}
\usepackage[compress,numbers]{natbib}
\usepackage{enumerate}
\usepackage[colorlinks=true,urlcolor=blue,anchorcolor=blue,citecolor=blue,filecolor=blue,linkcolor=blue,menucolor=blue,pagecolor=blue]{hyperref}

\allowdisplaybreaks

\makeatletter
\renewcommand{\@dotsep}{1000}
\makeatother

\newcommand{\captionfonts}{\small}
\makeatletter
\long\def\@makecaption#1#2{%
  \vskip\abovecaptionskip
  \sbox\@tempboxa{{\captionfonts #1: #2}}%
  \ifdim \wd\@tempboxa >\hsize
    {\captionfonts #1: #2\par}
  \else
    \hbox to\hsize{\hfil\box\@tempboxa\hfil}%
  \fi
  \vskip\belowcaptionskip}
\makeatother

\newcommand{\Sla}[1]%
{\kern0.12em{\raise.15ex\hbox{$/$}\kern-.74em #1}}

\numberwithin{equation}{section}

\long\def\symbolfootnote[#1]#2{\begingroup%
\def\thefootnote{\fnsymbol{footnote}}\footnote[#1]{#2}\endgroup}

\newcommand{\be}{\begin{equation}}
\newcommand{\ee}{\end{equation}}

\newcommand{\bea}{\begin{eqnarray}}
\newcommand{\eea}{\end{eqnarray}}
\newcommand{\mat}{\begin{pmatrix}}
\newcommand{\rix}{\end{pmatrix}}

\renewcommand{\bar}{\overline}
\renewcommand{\slash}[1]{#1\!\!\!/}

\newcommand{\go}{{\tilde g}}
\newcommand{\Bo}{{\tilde B}}
\newcommand{\Wo}{{\tilde W}}
\newcommand{\Ho}{{\tilde H}}

\newcommand{\cho}{{\tilde \chi}}
\newcommand{\sq}{{\tilde q}}
\newcommand{\st}{{\tilde t}}

\newcommand{\snu}{{\tilde\nu}}
\newcommand{\stau}{{\tilde\tau}}

\newcommand{\qq}{\qquad}

\newcommand{\beqa}{\begin{eqnarray}}
\newcommand{\eeqa}{\end{eqnarray}}

\newcommand{\beq}{\begin{equation}}
\newcommand{\eeq}{\end{equation}}
\newcommand{\hc}{\mbox{h.c.}}

\newcommand{\abs}[1]{\left\vert#1\right\vert}

\newcommand{\MET}{{\slash E_T}}

\newcommand{\jets}{\{\mbox{jets}\}}

\newcommand{\chk}{{\bf \surd}}
\newcommand{\xx}{{\bf \times}}

\renewcommand{\comment}[1]{}

\begin{document}

\begin{center}
{\Large\bf Sensitivity of an Upgraded LHC to  \\ R-Parity Violating Signatures of the MSSM }
\vspace{0.5cm}

Daniel Duggan$^\dagger$, \ \ 
Jared A. Evans$^\dagger$, \ \ 
James Hirschauer$^*$, \\ 
Ketino Kaadze$^*$,\ \ 
David Kolchmeyer$^\dagger$, \ \ 
Amit Lath$^\dagger$, \ \ 
Matthew Walker$^\dagger$ 
\vspace{0.6cm}

$^\dagger${\it Department of Physics and Astronomy, Rutgers University, Piscataway, NJ 08854}\\
$^*${\it Fermi National Accelerator Laboratory, Batavia, IL 60510}  \\
\end{center}

\begin{abstract}
We present a sensitivity study for the pair-production of supersymmetric particles which decay through R-parity violating channels.  As the scope of possible RPV signatures is very broad, the reach of several selected signatures spanning a representative variety of possible final states is considered.  Preference in representation is given to spectra motivated by naturalness, i.e. light higgsinos, stops and gluinos.  The sensitivity studies are presented for proton-proton collisions at 14 TeV with an integrated luminosity of 300 and 3000 fb$^{-1}$, as well as at 33 TeV with an integrated luminosity of 3000 fb$^{-1}$.\\
\end{abstract}

\section{Introduction}

The LHC has proven extremely effective at probing MSSM scenarios with conserved R-parity.  However, evidence for SUSY has not been found yet~\cite{ATLAS-SUSY-URL, CMS-SUSY-URL}, and the impressive experimental constraints are forcing the MSSM toward regions of parameter space unnatural for the Higgs sector.  Allowing R-parity violation (RPV) in the decays of superpartners is one mechanism to preserve a natural Higgs sector by evading existing experimental constraints.  These collider signatures have been probed in many variations by both theoretical~\cite{Brust:2012uf, Evans:2012bf, Han:2012cu,Franceschini:2012za,Bhattacherjee:2013gr, Berger:2013sir} and experimental studies~\cite{CMS-PAS-SUS-13-010, ATLAS-CONF-2013-036, Chatrchyan:2013xsw, Chatrchyan:2012sv, Chatrchyan:2011cj, ATLAS:2012dp, Chatrchyan:2013izb}. 

In this white paper for the 2013 Snomwmass study, an array of possible collider signatures from RPV are analyzed in the context of future LHC upgrade possibilities, particularly 14 TeV with integrated luminosities of 300 and 3000 fb$^{-1}$, as well as at 33 TeV with an integrated luminosity of 3000 fb$^{-1}$.  Although single superpartner production is, in principle, possible in RPV, these signatures tend to live on the fringe between what is viable and excluded by indirect bounds.  For this reason, we focus this work on the more generic signatures arising from pair-produced superpartners.  Each choice of production and decay mode constructs a separate simplified model~\cite{Alves:2011wf} which could be examined at the LHC.

The particular scenarios considered in the work are selected for several reasons.  First, the chosen final states serve to sample a diverse variety of the available options in RPV, covering many possibilities from multi-leptons to jet-only final states.  Second, many of the chosen options exhibit naturalness-motivated topologies, here defined to mean decays involving stops and higgsinos~\cite{Papucci:2011wy}.  Third, many of the chosen RPV couplings are compatible with those that might arise in the context of an RPV scenario exhibiting a ``third-generation dominant'' structure to the couplings, such as MFV SUSY \cite{Nikolidakis:2007fc, Csaki:2011ge, Krnjaic:2012aj, Franceschini:2013ne, Csaki:2013we}.

The $R$-parity violating superpotential and soft bilinear Lagrangian extension to the MSSM are:
\beqa
 W_{\rm RPV} &=& \frac12\lambda_{ijk} L_iL_j E^c_k + \lambda_{ijk}' L_iQ_j D^c_k + \frac12\lambda_{ijk}'' U^c_iD^c_j D^c_k + \mu_i L_i H_u \\
\mathcal L_{\rm soft\;RPV} &=& B_i \tilde{L}_i h_u + {\tilde m}^2_{di} h_d^\dagger \tilde{L}_i  +\hc \label{eq:RPVsoft}
\eeqa
In this study, we will refer to the trilinear couplings in the RPV superpotential, $\lambda_{ijk}$, $\lambda'_{ijk}$, and $\lambda''_{ijk}$, as $LLEijk$, $LQDijk$, $UDDijk$ respectively.  The RPV bilinear terms in the superpotential and soft lagrangian together allow for the neutral higgsino to decay as  $\Ho^0\to W^\pm \tau^\mp$.  The benchmark for bilinear RPV will be expressed as $LH3$.

As the strengths of individual couplings are unknown, but often highly constrained \cite{Barbier:2004ez, Kao:2009fg}, we will utilize a single coupling dominance ansatz to ameliorate issues from indirect constraints.  However, as the coupling strength does not generally contribute to pair production cross-sections, the specific value is largely irrelevant except insofar as all RPV decays are assumed to be both prompt and narrower than the detector resolution.   

The list of all benchmarks considered in this white paper is provided in Table~\ref{tab:rpv_benchmark}.  The next section of this work contains a catalogue of searches and the limits they set on the various simplified models arising from the RPV MSSM.  These searches are: multi-leptons, third-generation leptoquark, $\ell\;+$ many jets and paired dijets.  A summary of the projected 2$\sigma$ mass reaches is provided at the end of this white paper.

\begin{table}[t!]
\begin{center}
\begin{tabular}{|c|c|c|c|c|c|}
\hline
 Coupling & Production & Final States & Search & Natural & 3rd Gen \\
\hline\hline
$LLE122$ & $\go/\tilde{q}\to \Bo$ & $jj + \ell^+\ell^-\mu^+\mu^- +\MET$ & Multi-$\ell$ & $\xx$ & $\xx$ \\
                    & $\Wo$ & $\ell^+\ell^-\mu^+\mu^- +\MET$ & Multi-$\ell$& $\xx$ & $\xx$ \\
$LLE233$ & $\st \to \Ho$ & $b \bar b \tau^+ \tau^- \ell^+\ell^-+\MET$ & Multi-$\ell$& $\chk$ & $\chk$ \\
                    & $\Ho$ & $\tau^+ \tau^- \ell^+\ell^-+\MET$  & Multi-$\ell$ &$\chk$ & $\chk$ \\
\hline
$LQD232$ & $\go\to\st$ & $t \bar t \{\mu^+j\}  \{\mu^-j\}$ & Multi-$\ell$ & $\xx$ & $\xx$ \\
$LQD333$ & $\st$ & $\{\tau^+ b\}\{\tau^- b\}$ & 3rd Gen LQ &  $\chk$ & $\chk$ \\
\hline
$UDD212$ & $\st\to\Bo$ & $t\bar t \{jjj\}\{jjj\}$ & $\ell$  + $n$ jets & $\chk$ & $\xx$ \\
$UDD312$ & $\st$ & $\{jj\}\{jj\}$ & Dijet Pairs & $\chk$ & $\xx$\\
\hline
$LH3$ & $\Ho$ & $W^+W^-\tau^+\tau^-$ & Multi-$\ell$ & $\chk$ & $\chk$ \\
\hline
\end{tabular}
\caption{\label{tab:rpv_benchmark} Benchmark scenarios of RPV SUSY used in the Snowmass study.  The RPV coupling as well as production and RPC decays are presented in the first two columns.  The final particle in the chain is assumed to be the LSP for simplicity.  A typical final state arising from the cascade is presented in the third column, while the search strategy to address the signature is listed in the fourth columns.  Processes containing stops and higgsinos are labeled as ``Natural,'' while those possessing couplings suggestive of a third-generation dominance ansatz are labelled as ``3rd Gen.''   Further detail of the individual states is provided in the text.  A similar table (Table~\ref{tab:rpv_limits}) is provided in the conclusion where the projected sensitivities are tabulated. }
\end{center}
\end{table}

\section{Sensitivity Studies for the RPV Models \label{sec:sensitivitystudies}}

For all studies, Pythia 8~\cite{PY8} was used to generate the signal events.  These were processed using the detector simulator Delphes 3.0.6 or 3.0.9~\cite{deFavereau:2013fsa}.  The common Snowmass 2013 Energy Frontier Standard Model background samples \cite{Avetisyan:2013dta, Avetisyan:2013onh} were used in the majority of studies, except for in the paired dijet study of section \ref{sec:dij}.  Cuts were applied as detailed in each subsection.

\subsection{Multilepton studies  \label{sec:ML}}

\begin{figure}[t]
\begin{center}
\includegraphics[scale=.44]{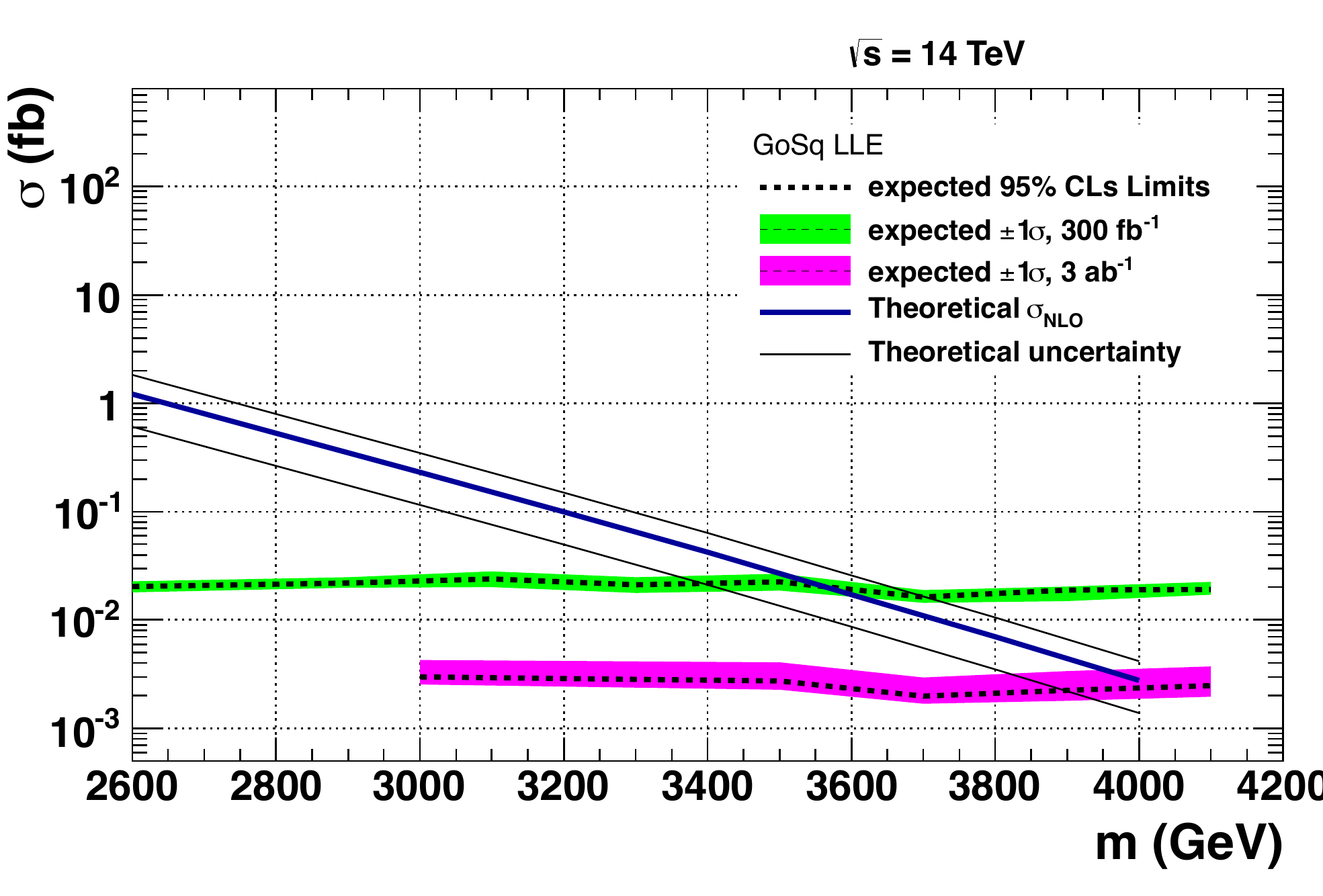}
\includegraphics[scale=.44]{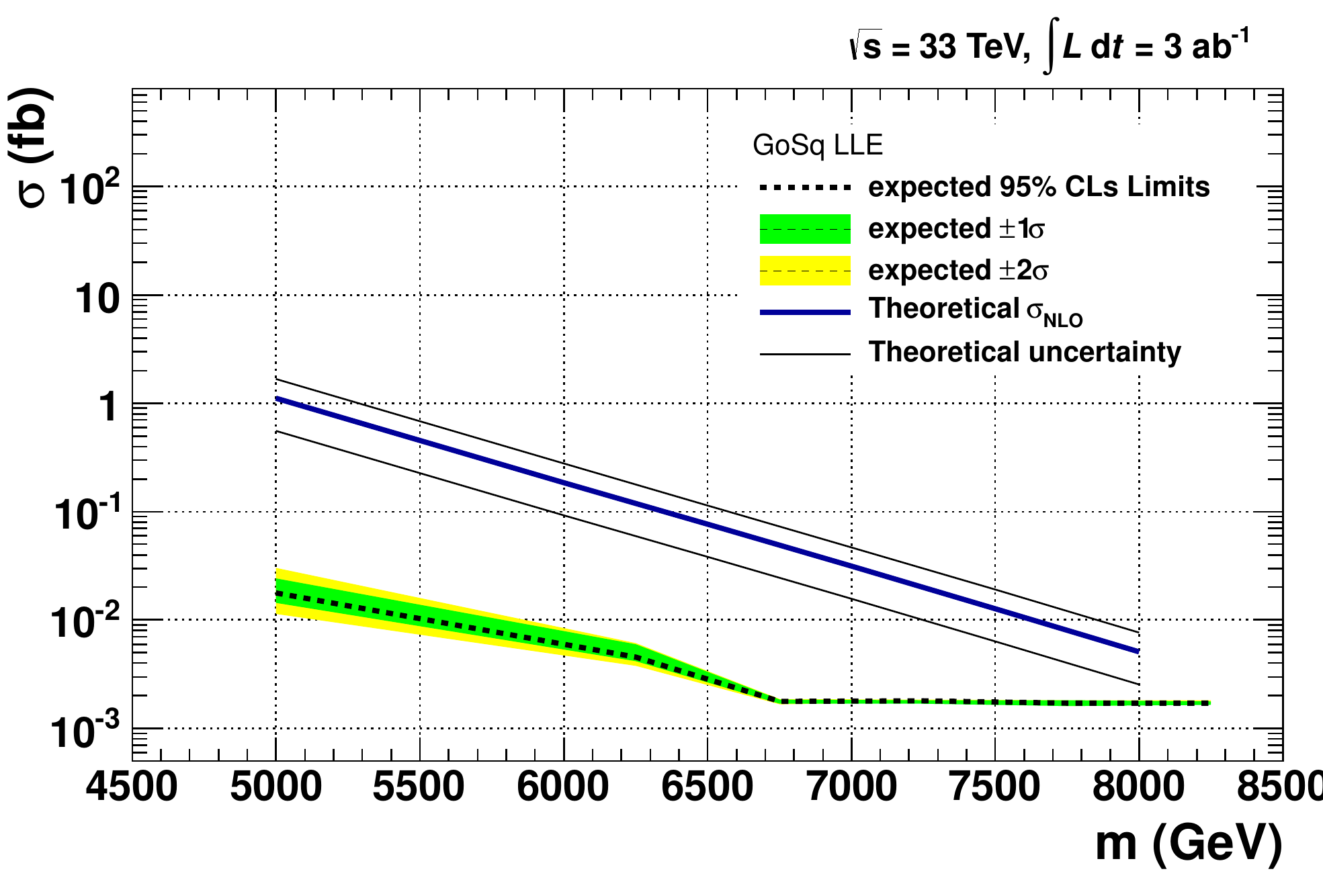}
\caption{Projected sensitivity from multi-lepton studies on production of degenerate squarks and gluinos decaying to jets and a neutralino which subsequently decays through the LLE122 operator. {\bf Left:} At 14 TeV with 300 fb$^{-1}$ and 3 ab$^{-1}$, the LHC has projected sensitivity to 3.55 TeV and 4.0 TeV respectively.  {\bf Right:} At 33 TeV with 3 ab$^{-1}$, the LHC has projected sensitivity to 8.5 TeV.  See section~\ref{sec:ML} for details on the model, selections and search regions used in the sensitivity study.}
\label{fig:MLGoSq}
\end{center}
\end{figure}

Multilepton searches are extremely powerful probes of new physics.  All models with RPV LLE couplings and with bilinear RPV can populate these searches.  Additionally, some models with LQD couplings, particularly those with tops, can give rise to multilepton signatures.  For this study, six distinct RPV simplified models are examined, however only a single search strategy is used following the framework of CMS multilepton studies~\cite{Chatrchyan:2013xsw}.  The events are binned exclusively across several variables: total leptons (3, 4+); $S_T$ (0-300, 300-600, 600-1000, 1000-1500, 1500-2000, 2000+); $b$-tags (0, 1, 2+);  $\tau$-tags (0, 1+); opposite-sign same-flavor pairs (0, 1, 2); and $Z$ candidates (0, 1+).  A combined fit is used to set exclusions.
\begin{itemize}
\item LLE122:  $\go/\tilde{q}\to \Bo$  --- Through various production mechanisms, such as $pp \to \tilde u \tilde u, \; \tilde u\go,\; \tilde u \tilde d,$ and $\tilde d \tilde d^*$,  first-generation squarks and gluinos are produced which then promptly decay to $j\Bo$ or $jj\Bo$, respectively.  The $\Bo$ then undergoes decays through an off-shell slepton and the LLE122 coupling to give an $e$- and $\mu$-rich final state, either $e^\pm\mu^\pm$ or $\mu^+\mu^-$ and a $\nu$ or $\bar\nu$.  This case has extremely high $S_T$ ($\sim2 m_\go$), significant $\MET$ from the neutrinos and large $H_T$ from the jets.  We use $m_\go : m_\sq : m_\Bo$ in a $2:2:1$ hierarchy for our choice of mass assignment (all other superpartners are decoupled).  Deviations from degenerate gluinos and squarks can change the production cross-section significantly, while changes to the bino mass would likely have a very small effect.  The LHC can probe extremely high masses in this benchmark allowing for incredible sensitivity.  As displayed in Figure~\ref{fig:MLGoSq}, LHC 14 with 300 fb$^{-1}$ (14 with 3 ab$^{-1}$ [33 with 3 ab$^{-1}$]) is expected to have sensitivity to $\go/\sq$ masses at 3.55 TeV (4.0 TeV [8.5 TeV]).
\begin{figure}[t]
\begin{center}
\includegraphics[scale=.44]{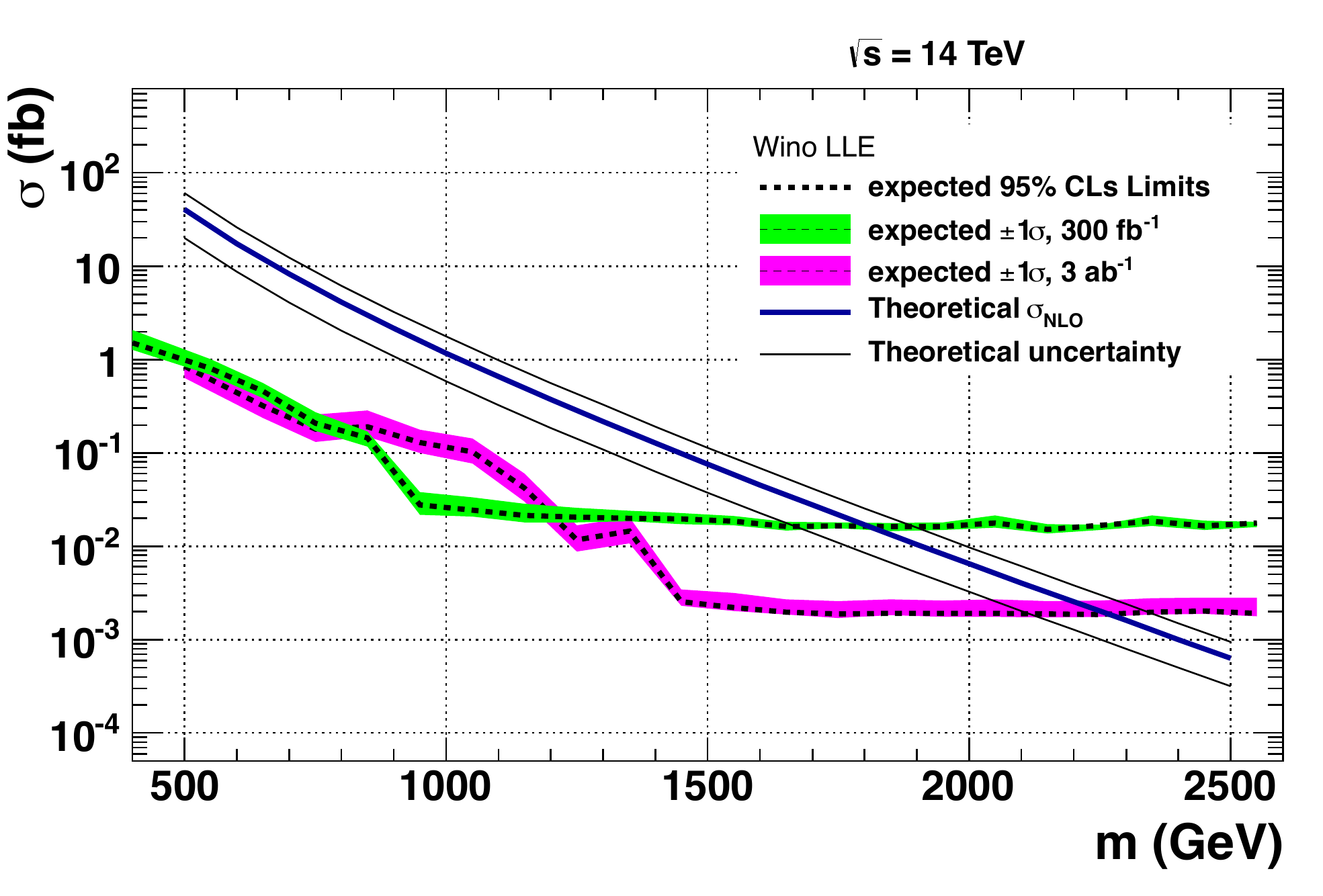}
\includegraphics[scale=.44]{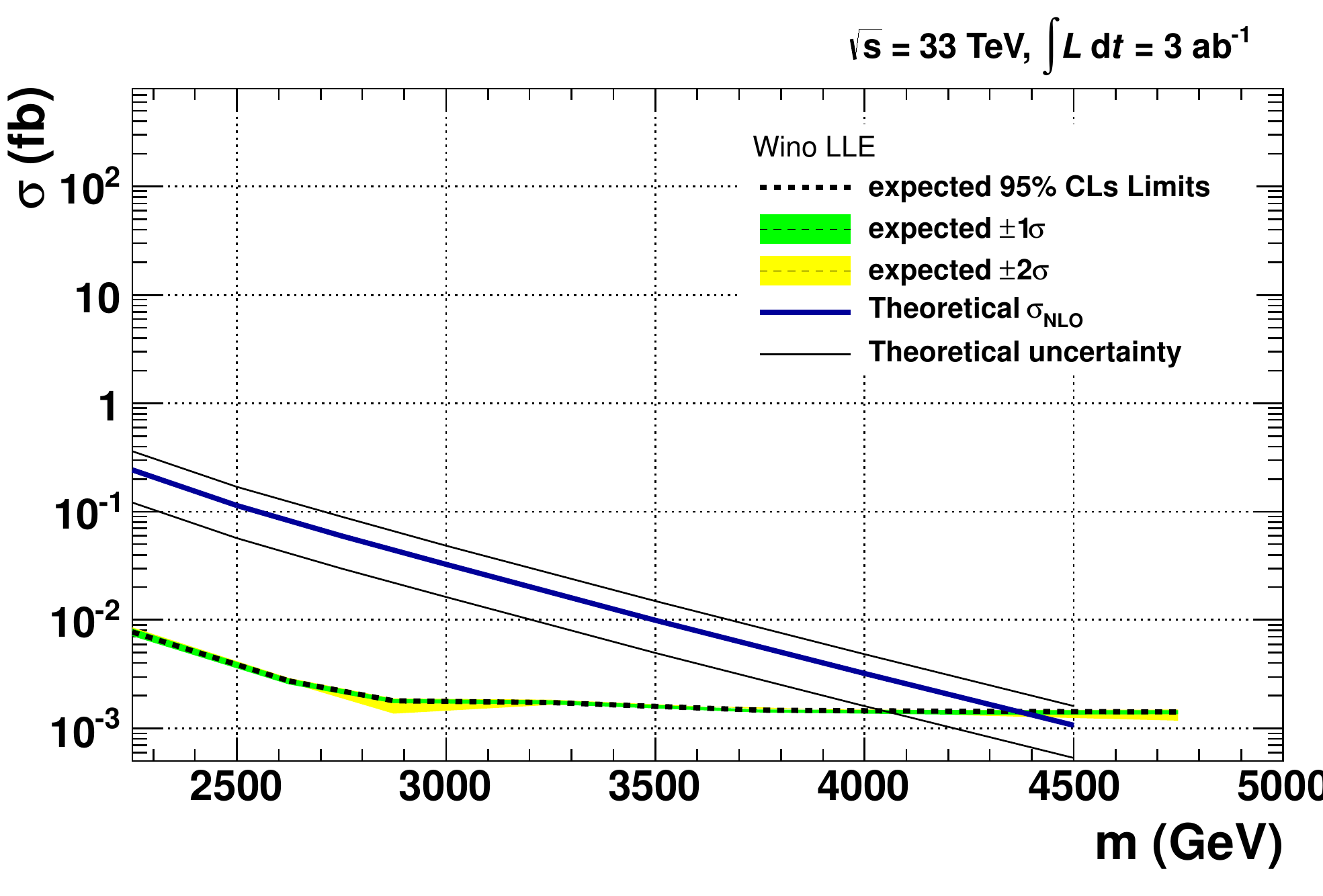}
\caption{Projected sensitivity from multi-lepton studies on production of winos which decay through the LLE122 operator. {\bf Left:} At 14 TeV with 300 fb$^{-1}$ and 3 ab$^{-1}$, the LHC has projected sensitivity to 1.8 TeV and 2.3 TeV respectively.  {\bf Right:} At 33 TeV with 3 ab$^{-1}$, the LHC has projected sensitivity to 4.4 TeV.  See section~\ref{sec:ML} for details on the model, selections and search regions used in the sensitivity study.}
\label{fig:MLWo}
\end{center}
\end{figure}
\item LLE122:  $\Wo$  --- In this benchmark model, charged and neutral winos are produced through an off-shell $W^\pm$ or $Z/\gamma$.  The charginos then undergo a prompt decay $\cho^\pm \to \cho^0 (W^{\pm*})$, where the off-shell $W$ is soft enough to be unobservable.  The wino-like $\cho^0$ then decays through an off-shell slepton to give an $e$- and $\mu$-rich final state, either $e^\pm\mu^\pm$ or $\mu^+\mu^-$ and a $\nu$ or $\bar\nu$.  This case has significant $\MET$ and $S_T\sim 2 m_\Wo$.  For this benchmark simplified model, the $\Wo$ appears alone at the bottom of the spectrum (all other superpartners are decoupled).  As displayed in Figure~\ref{fig:MLWo}, LHC 14 with 300 fb$^{-1}$ (14 with 3 ab$^{-1}$ [33 with 3 ab$^{-1}$]) is expected to have sensitivity to $\Wo$ masses at 1.8 TeV (2.3 TeV [4.4 TeV]).
\begin{figure}[t]
\begin{center}
\includegraphics[scale=.44]{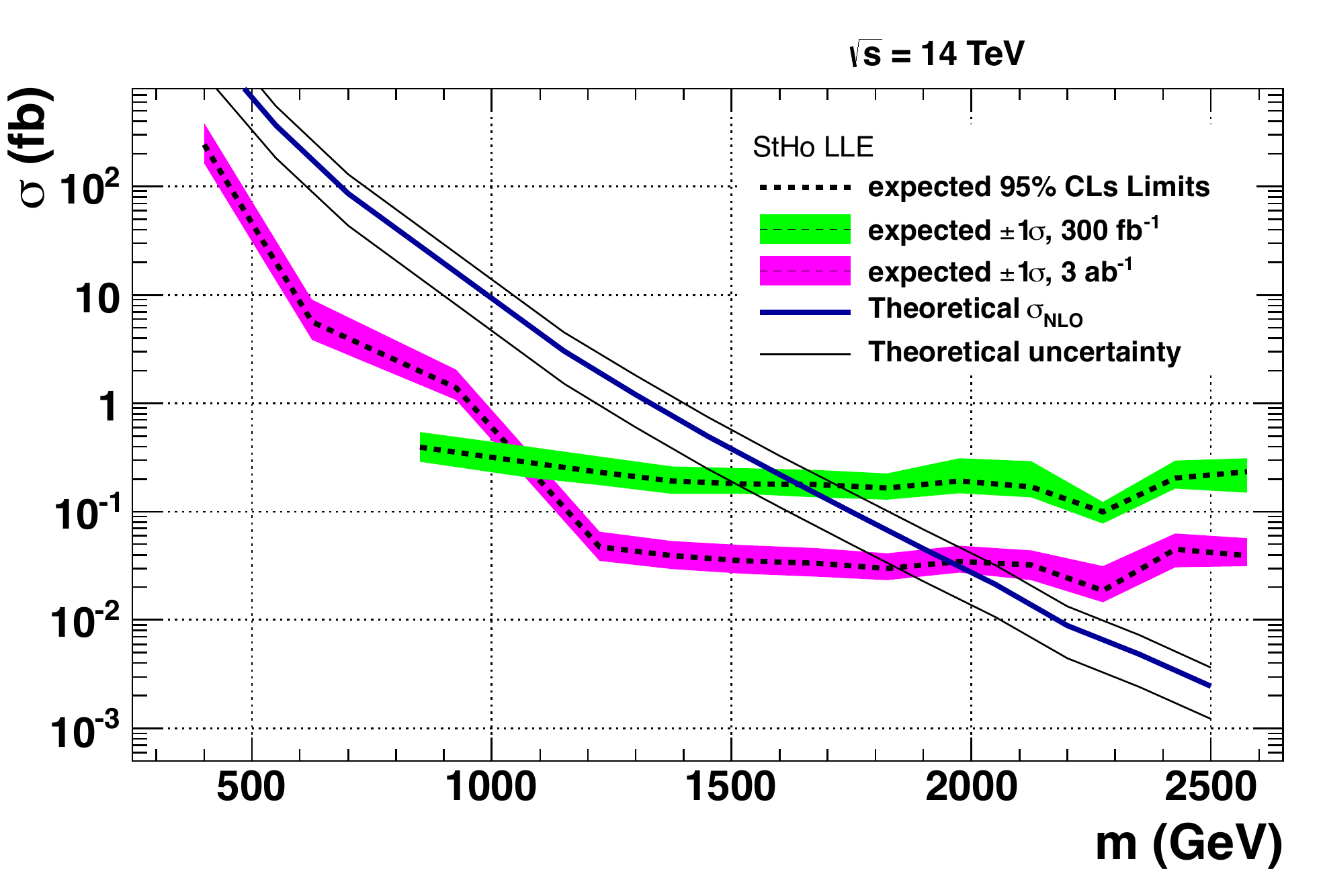}
\includegraphics[scale=.44]{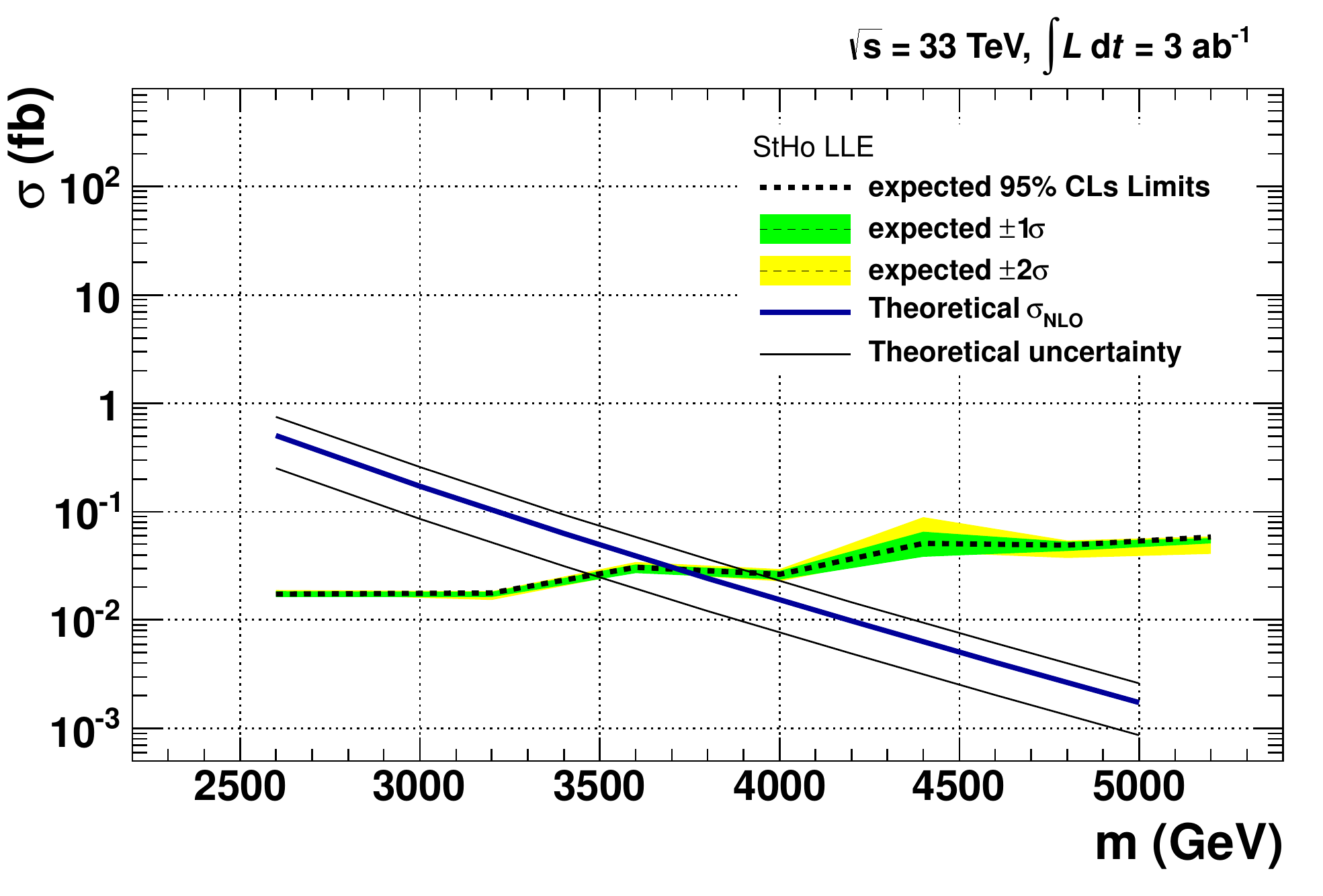}
\caption{Projected sensitivity from multi-lepton studies on production of stops which decay into charginos which then decay through the LLE233 operator. {\bf Left:} At 14 TeV with 300 fb$^{-1}$ and 3 ab$^{-1}$, the LHC has projected sensitivity to 1.65 TeV and 1.95 TeV respectively.  {\bf Right:} At 33 TeV with 3 ab$^{-1}$, the LHC has projected sensitivity to 3.75 TeV.  See section~\ref{sec:ML} for details on the model, selections and search regions used in the sensitivity study.}
\label{fig:MLStHoLLE}
\end{center}
\end{figure}
\item LLE233:  $\tilde{t}\to \Ho$  --- In this benchmark, pair-produced stop squarks decay promptly to $\Ho^\pm b$.  The $\Ho^\pm$ then promptly decays through an off-shell $\stau/\snu_\tau$ to give a $\tau$- and $\mu$-rich final state.  This case has significant $\MET$ and $S_T\sim 2m_\st$.  The masses are set so $m_\st - m_\Ho = 100$ GeV (all other states are decoupled) to forbid the $\st\to t \Ho^0$ decay.  The spectrum containing a stop and higgsino is motived by naturalness, while the coupling choice is compatible with a third-generation dominant scenario.   As displayed in Figure~\ref{fig:MLStHoLLE}, LHC 14 with 300 fb$^{-1}$ (14 with 3 ab$^{-1}$ [33 with 3 ab$^{-1}$]) is expected to have sensitivity to $\st$ masses at 1650 GeV (1950 GeV [3750 GeV]).
\begin{figure}[t]
\begin{center}
\includegraphics[scale=.44]{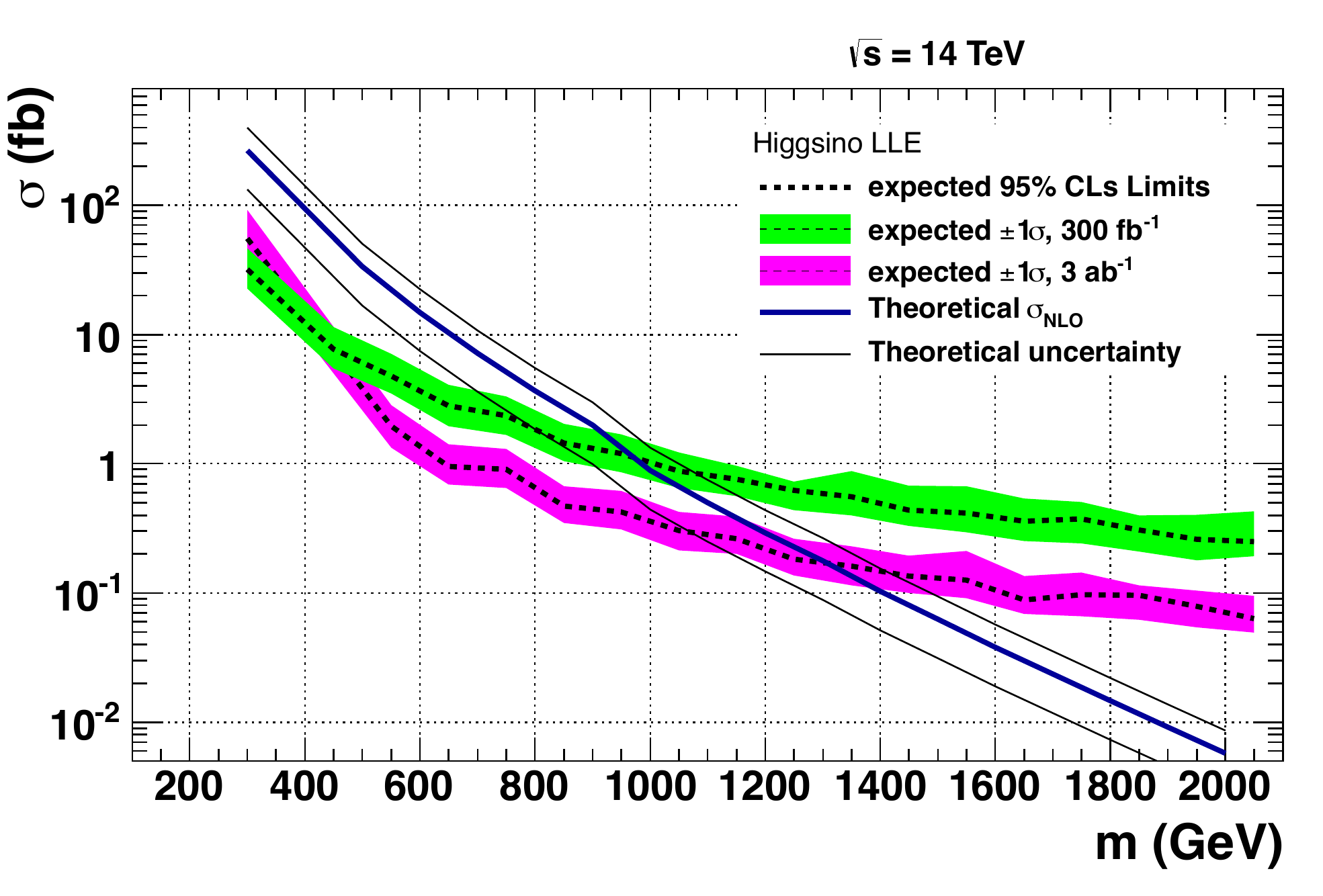}
\includegraphics[scale=.44]{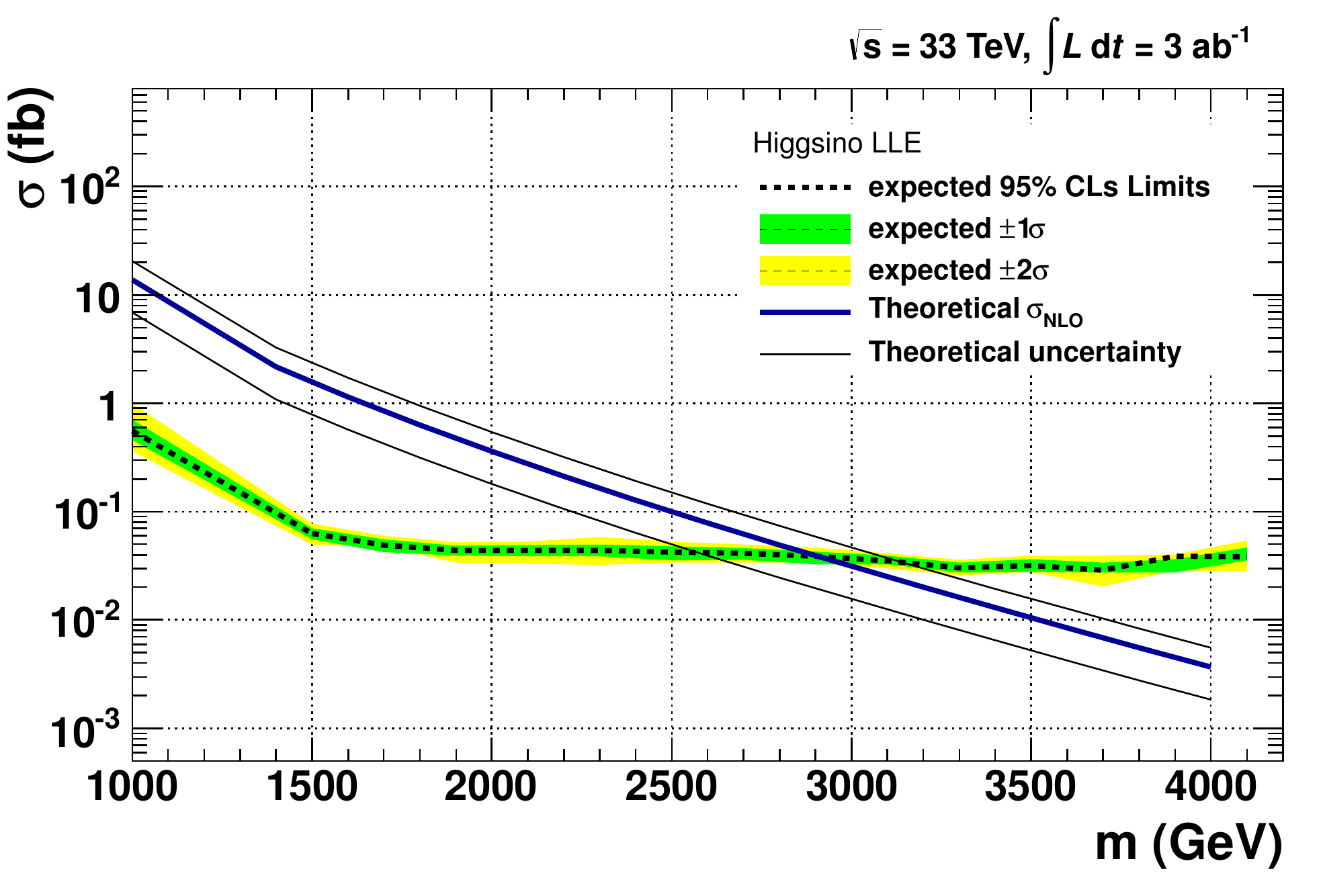}
\caption{Projected sensitivity from multi-lepton studies on production of higgsinos which then decay through the LLE233 operator. {\bf Left:} At 14 TeV with 300 fb$^{-1}$ and 3 ab$^{-1}$, the LHC has projected sensitivity to 950 GeV and 1300 GeV respectively.  {\bf Right:} At 33 TeV with 3 ab$^{-1}$, the LHC has projected sensitivity to 2.9 TeV.  See section~\ref{sec:ML} for details on the model, selections and search regions used in the sensitivity study.}
\label{fig:MLHoLLE}
\end{center}
\end{figure}
\item LLE233:  $\Ho$  --- In this benchmark, charged and neutral higgsino production are followed by prompt decays $\cho_2^0 \to \cho^0_1 (Z^*)$ or $ \cho^\pm (W^{\pm*})$ and $\cho^\pm \to \cho^0_1 (W^{\pm*})$, where the off-shell $W$s and $Z$s are soft enough to be unobservable.  The neutral $\Ho$ then decays through an off-shell $\stau/\snu_\tau$  to give a $\tau$- and $\mu$-rich final states.  This case has significant $\MET$.    A light higgsino is motived by naturalness, while the coupling is compatible with a third-generation dominant scenario.   As displayed in Figure~\ref{fig:MLHoLLE}, LHC 14 with 300 fb$^{-1}$ (14 with 3 ab$^{-1}$ [33 with 3 ab$^{-1}$]) is expected to have sensitivity to $\Ho$ masses at 950 GeV (1300 GeV [2900 GeV]).
\begin{figure}[t]
\begin{center}
\includegraphics[scale=.44]{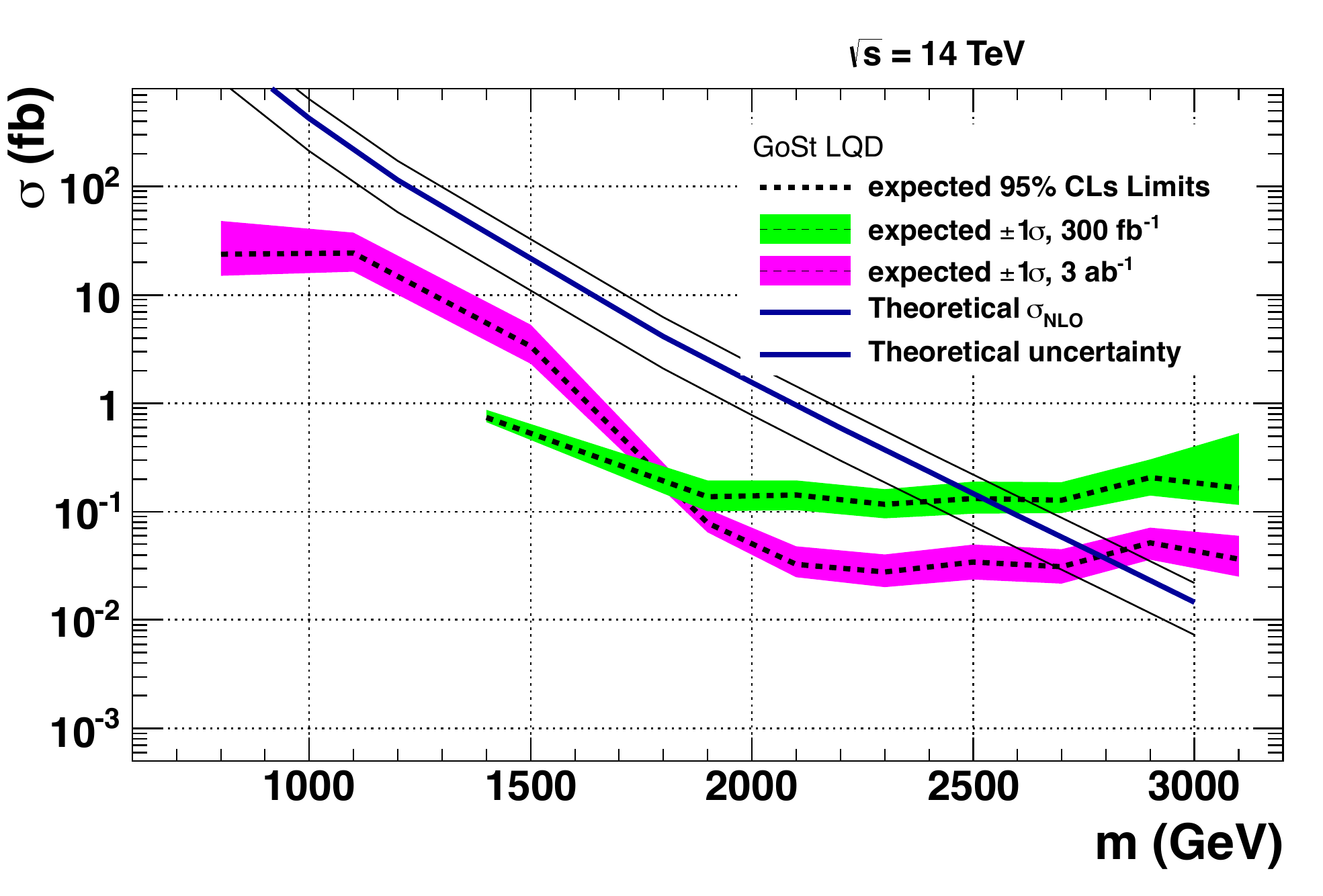}
\includegraphics[scale=.44]{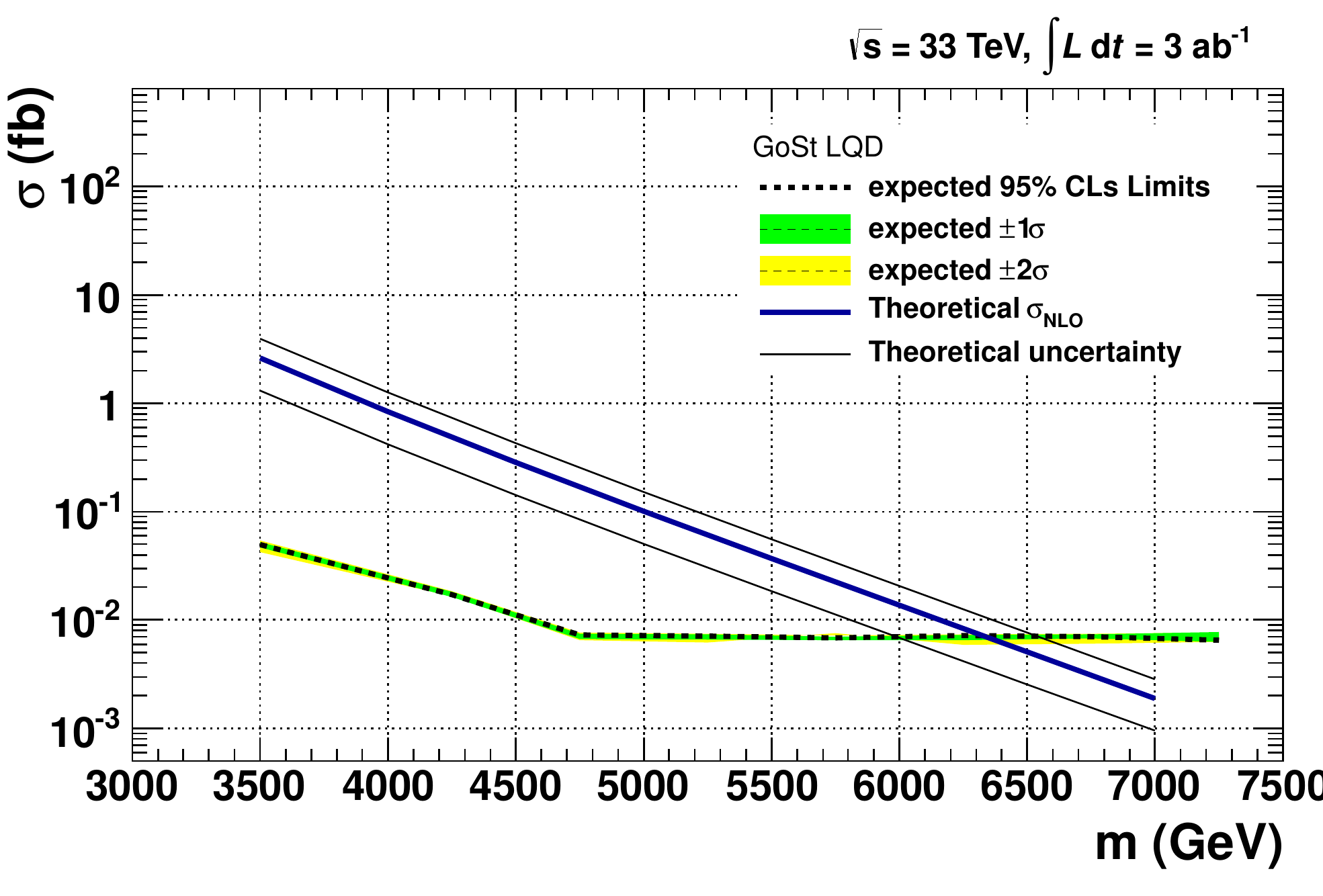}
\caption{Projected sensitivity from multi-lepton studies on production of gluinos which decay to stops which subsequently decay through the LQD232 operator.  {\bf Left:} At 14 TeV with 300 fb$^{-1}$ and 3 ab$^{-1}$, the LHC has projected sensitivity to 2.5 TeV and 2.8 TeV respectively.  {\bf Right:} At 33 TeV with 3 ab$^{-1}$, the LHC has projected sensitivity to 6.3 TeV.  See section~\ref{sec:ML} for details on the model, selections and search regions used in the sensitivity study.}
\label{fig:MLGoStLQD}
\end{center}
\end{figure}
\item LQD232:  $\go\to\st$  ---  Gluinos are pair-produced in this benchmark before decaying to $t\st$.  The stop then decays to a leptoquark-like final state ($\mu j$).  As a benchmark simplified model, we propose $m_\go - m_\st = 200$ GeV (with all other states decoupled).  Although a multilepton search is used to determine the sensitivity to this simplified model, same-sign dilepton or leptoquark resonance searches could have comparable sensitivity.   As displayed in Figure~\ref{fig:MLGoStLQD}, LHC 14 with 300 fb$^{-1}$ (14 with 3 ab$^{-1}$ [33 with 3 ab$^{-1}$]) is expected to have sensitivity to $\go$ masses at 2.5 TeV (2.8 TeV [6.3 TeV]).
\begin{figure}[t]
\begin{center}
\includegraphics[scale=.44]{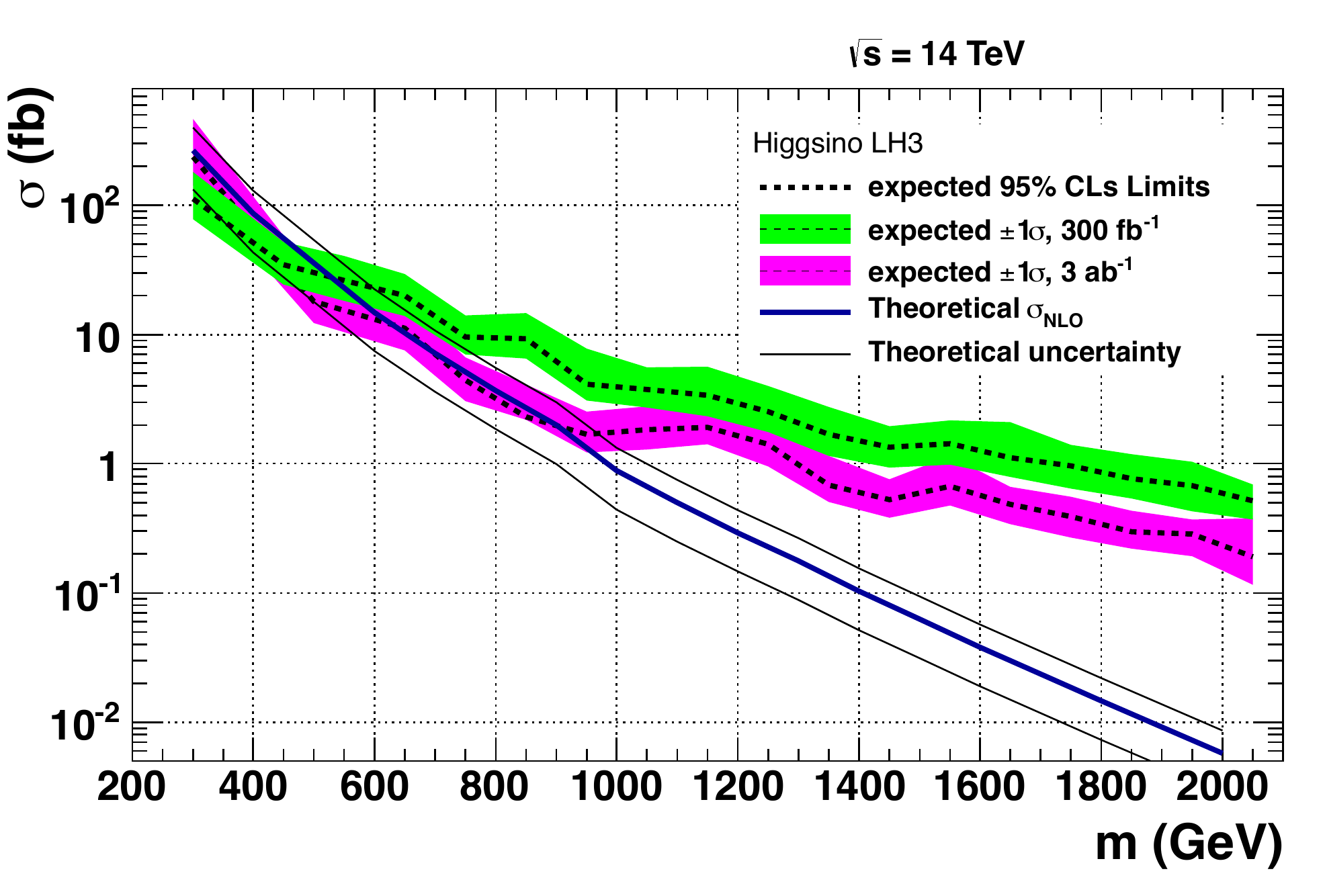}
\includegraphics[scale=.44]{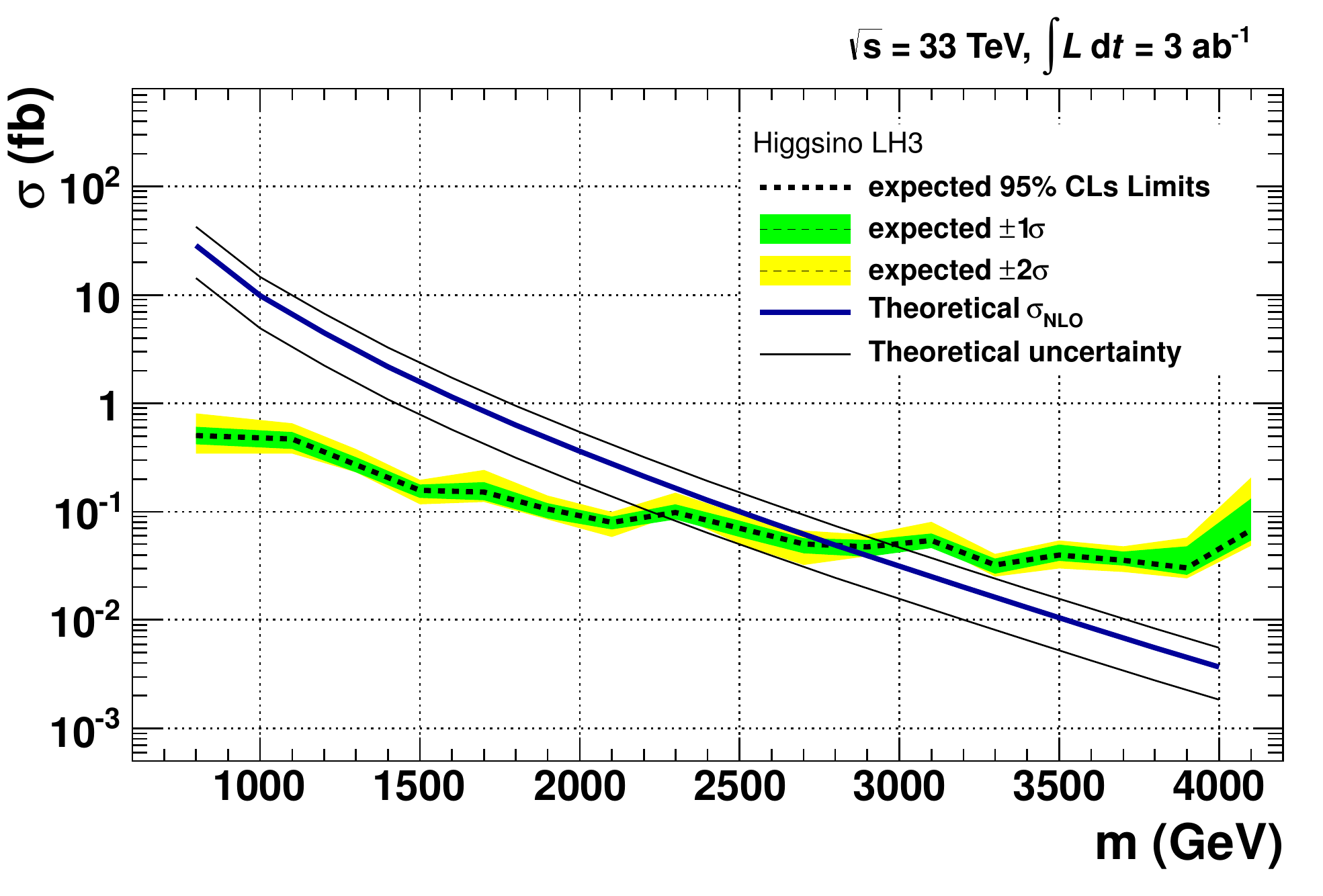}
\caption{Projected sensitivity from multi-lepton studies on production of higgsinos which then decay through the bilinear RPV LH3 operator to $W^\pm\tau^\mp$. {\bf Left:} At 14 TeV with 300 fb$^{-1}$ and 3 ab$^{-1}$, the LHC has projected sensitivity to 530 GeV and 610 GeV respectively.  {\bf Right:} At 33 TeV with 3 ab$^{-1}$, the LHC has projected sensitivity to 2800 TeV.  See section~\ref{sec:ML} for details on the model, selections and search regions used in the sensitivity study.}
\label{fig:MLHoLH}
\end{center}
\end{figure}
\item LH3:  $\Ho$  --- This benchmark has charged and neutral higgsino production with subsequent prompt decays $\cho_2^0 \to \cho^0_1 (Z^*)$ or $ \cho^\pm (W^{\pm*})$ and $\cho^\pm \to \cho^0_1 (W^{\pm*})$, followed by $ \cho^0_1 \to W^\pm \tau^\mp$.   As a benchmark simplified model, $\Ho$ appears alone at the bottom of the spectrum (with all other states decoupled).  The higgsino at the bottom of the spectrum is motivated by naturalness, while the LH3 coupling is compatible with a third-generation dominant ansatz.   As displayed in Figure~\ref{fig:MLHoLH}, LHC 14 with 300 fb$^{-1}$ (14 with 3 ab$^{-1}$ [33 with 3 ab$^{-1}$]) is expected to have sensitivity to $\Ho$ masses at 530 GeV (610 GeV [2800 GeV]).
\end{itemize}

\subsection{Third-generation leptoquark study \label{sec:LQ3}}

\begin{figure}[t]
\begin{center}
\includegraphics[scale=.42]{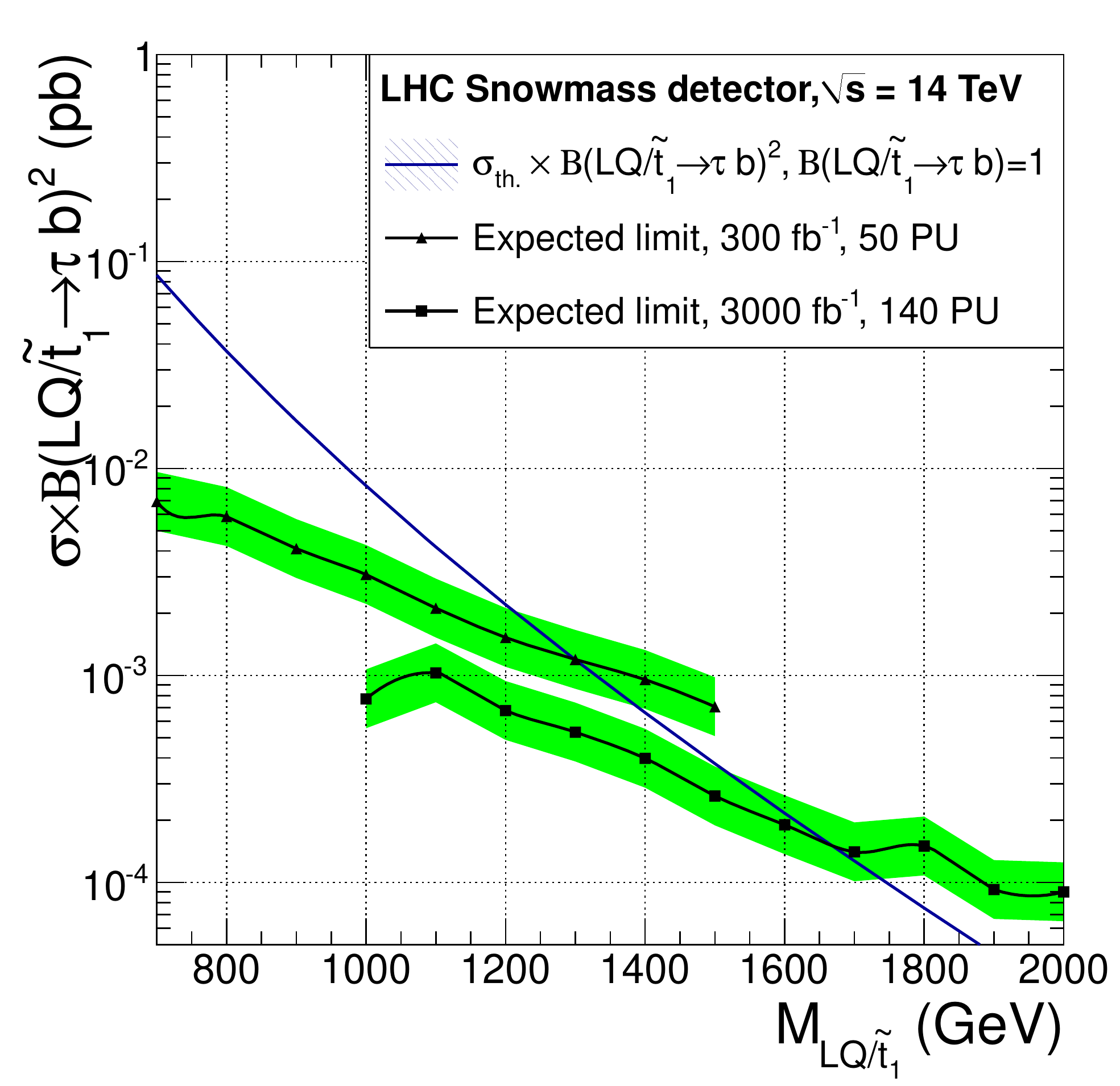} 
\caption{Limits on stops decaying to a $\tau b$ final state through the LQD333 operator.  This signature is identical to third-generation leptoquarks in terms of both pair production cross-section and decay modes.  For this study, only the $\mu \tau_h$ signature is used, the addition of electrons would increase sensitivity by a factor of $\sim 2$.    With 300 fb$^{-1}$, LHC 14 could exclude these stops (third-generation scalar leptoquarks) up to 1.3 TeV, while with 3 ab$^{-1}$ projected exclusion reaches 1.65 TeV.  See section~\ref{sec:LQ3} for details on the model, selections and search regions used in the sensitivity study. }
\label{fig:LQ3}
\end{center}
\end{figure}

R-parity violating stops can mimic leptoquarks (LQ).  Due to the enhanced motivation from third-generation dominance, the LQD333 operator is chosen to be studied as a benchmark. 

\begin{itemize}
\item LQD333:  $\st$  ---  Stops are pair-produced followed by a direct decay to $\tau b$.  This signal is identical to a third-generation leptoquark.  As a benchmark simplified model, we propose $m_\st$ is at the bottom of the spectrum with all other states decoupled.  A light stop is motived by naturalness and the coupling strength is suggestive of a third-generation dominant scenario.
\end{itemize}

For this study, the following selections are applied:  
\begin{itemize}
\item Require a $\mu$ with $p_T>30$ GeV and $\abs{\eta} < 2.1$
\item Require a $\tau_h$ with $p_T>50$ GeV, $\abs{\eta} < 2.3$ and a charge that is the opposite sign of the $\mu$ 
\item Require two $b$-tagged jets with $p_T>30$ GeV and $\abs{\eta} < 2.4$
\item All objects separated by $\Delta R\geq 0.5$ 
\end{itemize}
To reject the major backgrounds -- $t\bar t+\jets$, $Z+\jets$ and $W+\jets$ -- additional mass-dependent cuts are applied to $S_T$ and the invariant mass of the hadronic $\tau$ and $b$-jet:
\begin{itemize}
\item $S_T \equiv p_T(\mu)+p_T(\tau_h)+p_T(b_1)+p_T(b_2) > 1.25 m_\st$
\item $ M(\tau_h, b_i) >0.5 m_\st$, where $i\in\{1,2\}$ is the choice that minimizes $\abs{M(\tau_h, b_{i})-M(\mu_h, b_{\neq i})}$
\end{itemize}
For this study, only the $\mu \tau_h$ final state is used, but the inclusion of $e\tau_h$ could increase sensitivity by a factor of $\sim 2$.  For 3 ab$^{-1}$ (140 PU), the systematic uncertainties due to object identification or misidentification rates were inflated by 50\% from the nominal values.  As displayed in Figure~\ref{fig:LQ3}, LHC 14 with 300 fb$^{-1}$ (14 with 3 ab$^{-1}$) is expected to have sensitivity to $\st$ masses at 1.30 TeV (1.65 TeV).

\subsection{$\ell\;+$ many jets study \label{sec:lnj}}

\begin{figure}[t]
\begin{center}
\includegraphics[scale=.42]{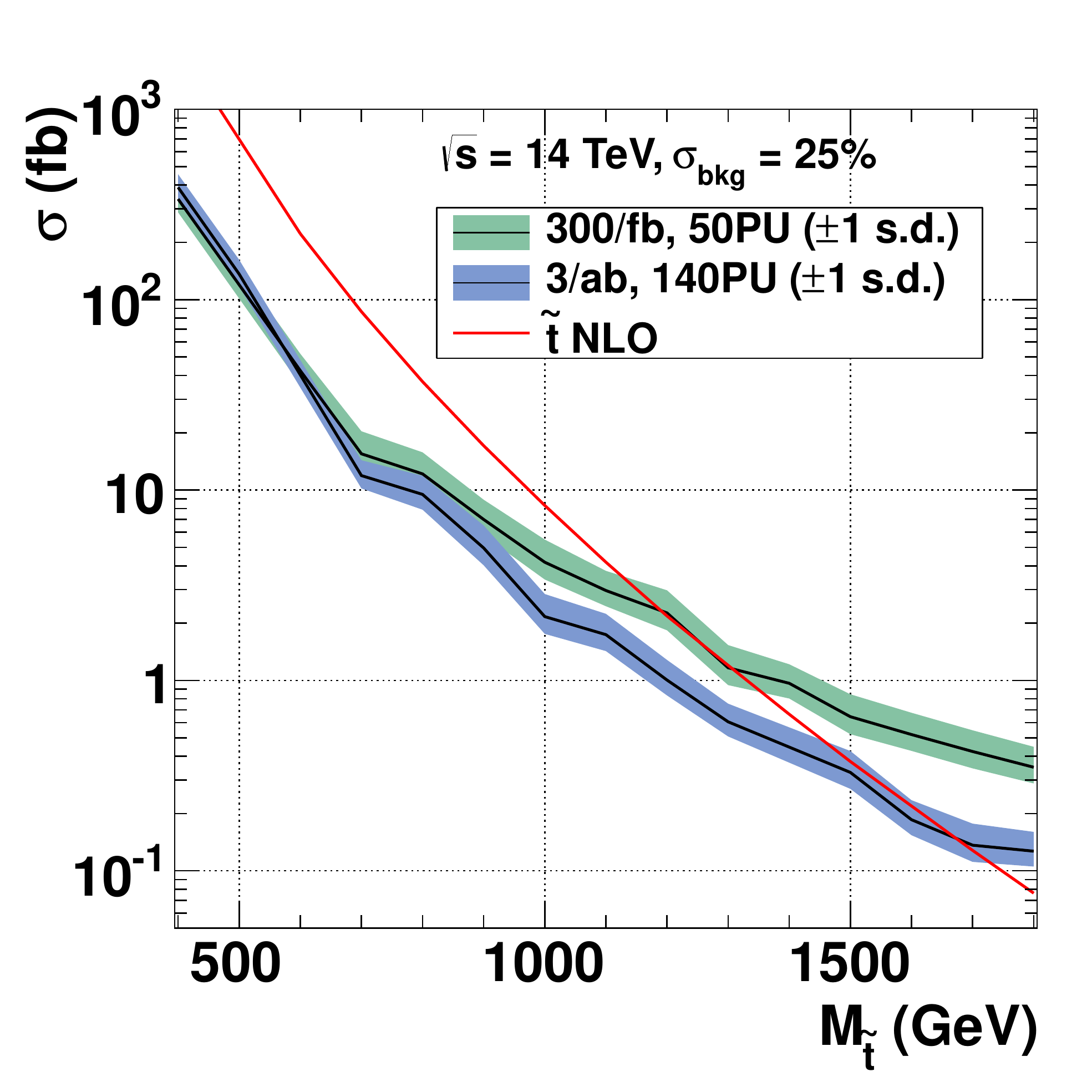} \qq
\includegraphics[scale=.42]{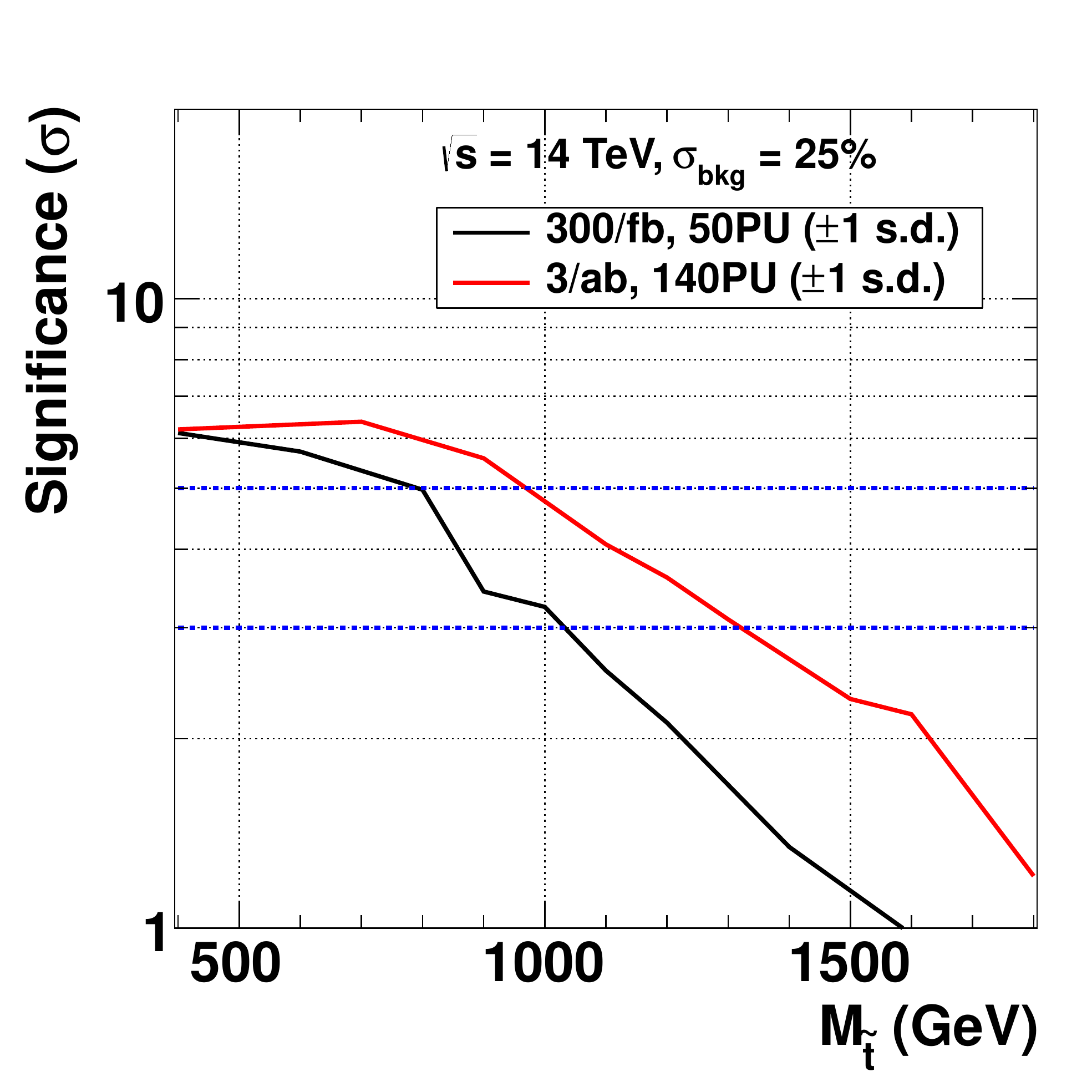} \qq
\caption{{\bf Left:}  Limits on stops decaying to a $t \cho^0 \to t \{jjj\}$ final state through the UDD212 operator.   With 300 fb$^{-1}$, LHC 14 could exclude these stops up to 1.2 TeV, while with 3 ab$^{-1}$ projected exclusion reaches 1.65 TeV. 
{\bf Right:}  Projected discovery significance for the same model.   See section~\ref{sec:lnj} for details on the model, selections and search regions used in the sensitivity study.}
\label{fig:lnj}
\end{center}
\end{figure}

The UDD RPV operators can easily give rise to very high jet multiplicities.  The possibility of tops or $W$s appearing in viable, simple decays motivates searches for $\ell+n$ jets.
\begin{itemize}
\item UDD212:  $\st\to \Bo$  --- In this benchmark, pair-produced stop squarks each decay to $t\Bo$.  The $\Bo$ subsequently decays then through an off-shell $\sq$ to give a three light jet resonance.   We use $m_\st : m_\Bo$ in a $2:1$ hierarchy with all other states decoupled.  The presence of a light stop is motived by naturalness.
\end{itemize}

For this study, the following selections are applied:  
\begin{itemize}
\item Require exactly one $e$ or $\mu$ with $p_T>30$ GeV and $\abs{\eta} < 2.1$
\item Require seven or more jets with $p_T>50$ GeV and $\abs{\eta} < 2.5$
\item Require at least one $b$-tagged jet with  $p_T>50$ GeV and $\abs{\eta} < 2.5$
\end{itemize}
To reject the major backgrounds -- $t\bar t+\jets$, $Z+\jets$ and $W+\jets$ -- a boosted decision tree (BDT) is used with $N_j$, $N_b$, $S_T$ and the $p_T$, $\phi$ and $\eta$ of the lepton and leading 6 jets.  In an analysis of real data, the BDT-based analysis would be extensively validated with a simpler, though slightly less potent, cut-based analysis.   As displayed in Figure~\ref{fig:lnj}, LHC 14 with 300 fb$^{-1}$ (14 with 3 ab$^{-1}$) is expected to have sensitivity to $\st$ masses at 1200 GeV (1650 GeV).

\subsection{Paired dijet study  \label{sec:dij}}

\begin{figure}[t]
\begin{center}
\includegraphics[scale=0.9]{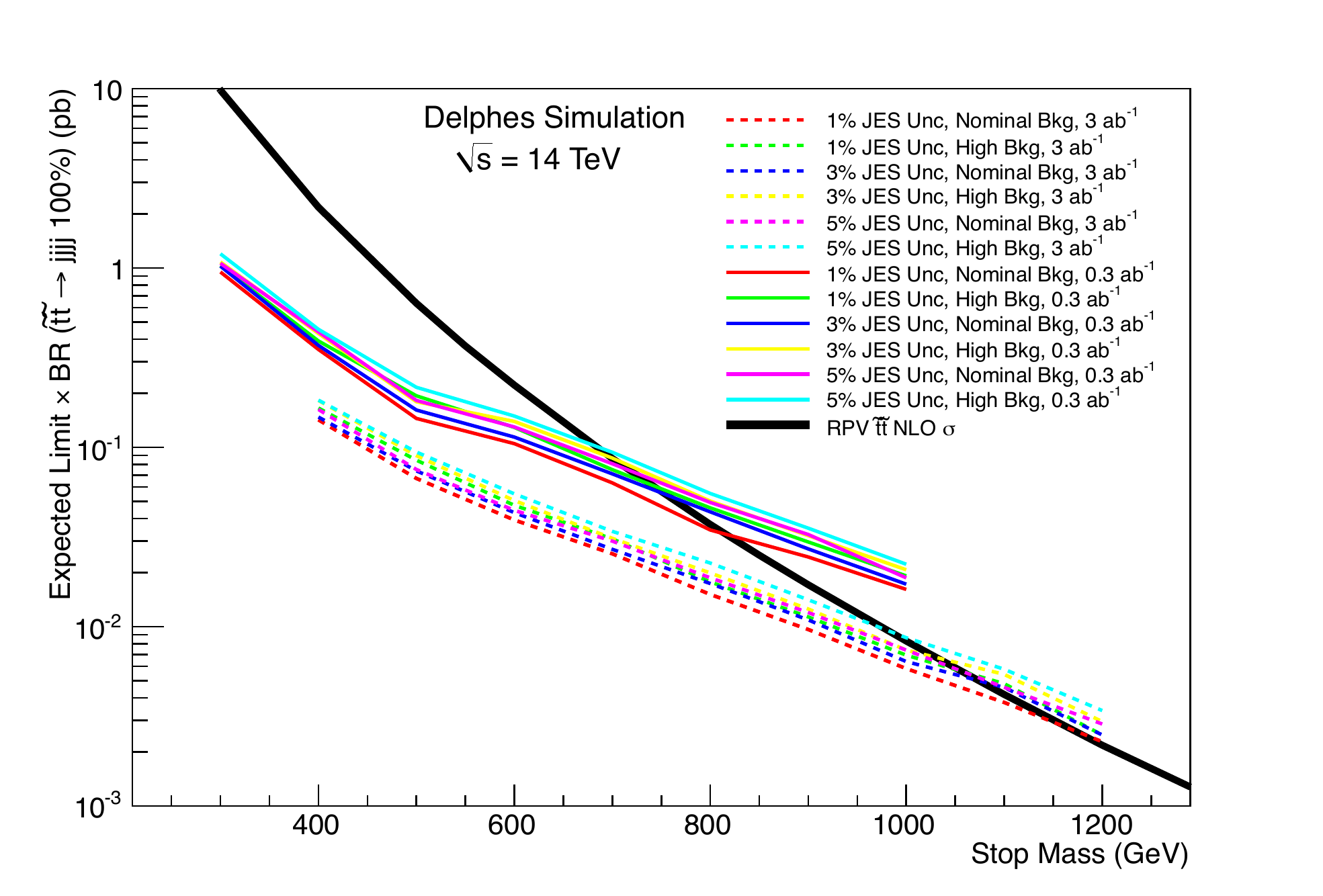}
\caption{Limits on pair-produced stops each decaying to a $jj$ final state through the UDD312 operator.   With 300 fb$^{-1}$, LHC 14 could exclude (conservatively : optimistically) these RPV stops up to 750  (680 : 840) GeV, while with 3 ab$^{-1}$ projected exclusion reaches 1070  (1000 : 1180) GeV.  See section~\ref{sec:dij} for details on the model, selections and search regions used in the sensitivity study.}
\label{fig:dij}
\end{center}
\end{figure}

A very interesting possibility for concealing supersymmetry with RPV  is the possibility of pair-produced stops decaying to a pair of dijet resonances.  
\begin{itemize}
\item UDD312:  $\st$  ---  For this benchmark, stops are pair-produced followed by a direct decay to $jj$, giving the signature of a pair of dijet resonances.   This model has the stop at the bottom of the spectrum with all other states decoupled.  A light stop is motived by naturalness.  This coupling is not motivated from third-generation dominance, however, this is a more difficult signature than the much easier $\{jb\}$ resonant pairs arising from the  third-generation dominant UDD323.
\end{itemize}
This study requires four or more jets with $p_T>120$ and $\abs{\eta} < 2.5$. From the four highest $p_T$ jets, all three unique combinations of paired dijets are composed, e.g.~the dijet composed from the first and second leading jets will be paired with the dijet constructed from the third and fourth leading jets. The combination resulting in the lowest fractional mass difference, i.e.~$\frac{|M_{a} - M_{b}|}{(M_{b} + M_{a})/2}$, is selected.  Additionally, we define $\Delta$ to be the difference between the scalar sum of the transverse momenta of the two jets in the dijet and the average pair mass in the event, e.g.~$\Delta = \sum_{1,2}|p_{T}^{i}| - (M_a+M_b)/2$.  As stops with greater boost would have higher $\Delta$, a cut of 50 GeV is implemented to efficiently remove the enormous QCD multi-jet background.  After all selection requirements have been satisfied, the average mass distribution is investigated for the sensitivity of signal in the presence of QCD background events.  

Because no monte carlo simulation is expected to be reliable for a background estimate, we instead employ a data-driven background scaling. In the CMS analysis~\cite{Chatrchyan:2013izb}, the background QCD distribution is modeled by a four-parameter function:
\beq
f(x) = P_0 \frac{(1 - x\/ \sqrt{s})^{P_1}}{(x\/ \sqrt{s})^{P_2 + P_3 \log{x\/ \sqrt{s}}}}
\eeq
Using the parameter values based on those from the fit to the 7 TeV data, an estimate of the background can be made simply by scaling up the normalization, $P_0$. The ratio between the $t\bar{t}$ cross-sections at 7 TeV and 14 TeV is taken as the scaling factor representing the nominal background expectation. A second, high background case is considered, quantified via increasing this ratio by 40\%.  At both the nominal and high backgrounds, three choices of jet-energy scale (JES) uncertainty (1\%, 3\% and 5\%) are used to further gauge the plausible variation.  As displayed in Figure~\ref{fig:dij}, LHC 14 with 300 fb$^{-1}$ (14 with 3 ab$^{-1}$) is expected to have sensitivity to $\st$ masses at 750 GeV (1070 GeV) with nominal backgrounds and a 3\% JES uncertainty.

\section{Summary}

\begin{table}[t!]
\begin{center}
\begin{tabular}{|c|c|c|c|c|c|c|}
\hline
 Coupling & Production & Final States & Search & 300 fb$^{-1}$ & 3 ab$^{-1}$ & 33 TeV \\
\hline\hline
$LLE122$ & $\go/\tilde{q}\to \Bo$ & $jj + \ell^+\ell^-\mu^+\mu^- +\MET$ & Multi-$\ell$ & 3550 & 4000 & 8500 \\
                    & $\Wo$ & $\ell^+\ell^-\mu^+\mu^- +\MET$ & Multi-$\ell$ & 1800 & 2300 &4400 \\
$LLE233$ & $\st \to \Ho$ & $b \bar b \tau^+ \tau^- \ell^+\ell^-+\MET$ & Multi-$\ell$& 1650 & 1950 & 3750 \\
                    & $\Ho$ & $\tau^+ \tau^- \ell^+\ell^-+\MET$  & Multi-$\ell$& 950 & 1300 & 2900 \\
\hline
$LQD232$ & $\go\to\st$ & $t \bar t \{\mu^+j\}  \{\mu^-j\}$ & Multi-$\ell$ & 2500 & 2800  & 6300 \\
$LQD333$ & $\st$ & $\{\tau^+ b\}\{\tau^- b\}$ & 3G LQ & 1300 & 1650 & --- \\
\hline
$UDD212$ & $\st\to\Bo$ & $t\bar t \{jjj\}\{jjj\}$ & $\ell$  + $n$ jets & 1200 & 1650 & --- \\
$UDD312$ & $\st$ & $\{jj\}\{jj\}$ & Dijet Pairs& 750 & 1070  & --- \\
\hline
$LH3$ & $\Ho$ & $W^+W^-\tau^+\tau^-$ & Multi-$\ell$ & 530 & 610  & 2800 \\
\hline
\end{tabular}
\caption{\label{tab:rpv_limits} Limits on the various benchmark scenarios of RPV SUSY considered in the Snowmass study.   Further detail of the individual states is provided in the text of section~\ref{sec:sensitivitystudies}.}
\end{center}
\end{table}

In this work, a representative variety of plausible collider signatures arising in R-parity violating supersymmetry were catalogued and studied.  These signatures were chosen as they sample the breadth of possibilities available in RPV SUSY.  Many of the signatures were selected due to their compatibility with a natural Higgs sector and/or a third-generation dominant ansatz for the RPV couplings.   At 14 TeV with both 300 fb$^{-1}$ and 3000 fb$^{-1}$ (represented by 50 pile-up events and 140 pile-up events respectively), sensitivity studies were performed gauging the reach of future colliders to these simplified models.  In some cases, 33 TeV with 3 ab$^{-1}$ was also studied.   These limits are summarized in Table~\ref{tab:rpv_limits}.

\section*{Acknowledgements}

We would like to thank the Snowmass conveners and organizers, particularly C.~Brock, Y.~Gershtein, M.~Luty, M.~Narain, M.~Peskin, L.~Wang and D.~Whiteson.  Additionally, JAE would like to thank Y.~Kats for useful discussions.

\small{\bibliography{RPV}}

\providecommand{\href}[2]{#2}\begingroup\raggedright\begin{thebibliography}{10}

\bibitem{ATLAS-SUSY-URL}
{ ATLAS} collaboration, ``{ATLAS Supersymmetry Searches},''
  \url{https://twiki.cern.ch/twiki/bin/view/AtlasPublic/SupersymmetryPublicRes%
ults} (2012).

\bibitem{CMS-SUSY-URL}
{ CMS} collaboration, ``{CMS Supersymmetry Physics Results},''
  \url{https://twiki.cern.ch/twiki/bin/view/CMSPublic/PhysicsResultsSUS}
  (2012).

\bibitem{Brust:2012uf}
C.~Brust, A.~Katz, and R.~Sundrum, ``{SUSY Stops at a Bump},''
  \href{http://dx.doi.org/10.1007/JHEP08(2012)059}{{\em JHEP} {\bfseries 1208}
  (2012) 059},
\href{http://arxiv.org/abs/1206.2353}{{\ttfamily arXiv:1206.2353 [hep-ph]}}.

\bibitem{Evans:2012bf}
J.~A. Evans and Y.~Kats, ``{LHC Coverage of RPV MSSM with Light Stops},''
  \href{http://dx.doi.org/10.1007/JHEP04(2013)028}{{\em JHEP} {\bfseries 1304}
  (2013) 028},
\href{http://arxiv.org/abs/1209.0764}{{\ttfamily arXiv:1209.0764 [hep-ph]}}.

\bibitem{Han:2012cu}
Z.~Han, A.~Katz, M.~Son, and B.~Tweedie, ``{Boosting Searches for Natural SUSY
  with RPV via Gluino Cascades},''
\href{http://arxiv.org/abs/1211.4025}{{\ttfamily arXiv:1211.4025 [hep-ph]}}.

\bibitem{Franceschini:2012za}
R.~Franceschini and R.~Torre, ``{RPV stops bump off the background},''
  \href{http://dx.doi.org/10.1140/epjc/s10052-013-2422-x}{{\em Eur.Phys.J.}
  {\bfseries C73} (2013) 2422},
\href{http://arxiv.org/abs/1212.3622}{{\ttfamily arXiv:1212.3622 [hep-ph]}}.

\bibitem{Bhattacherjee:2013gr}
B.~Bhattacherjee, J.~L. Evans, M.~Ibe, S.~Matsumoto, and T.~T. Yanagida,
  ``{Natural SUSY's Last Hope: R-parity Violation via UDD Operators},''
\href{http://arxiv.org/abs/1301.2336}{{\ttfamily arXiv:1301.2336 [hep-ph]}}.

\bibitem{Berger:2013sir}
J.~Berger, M.~Perelstein, M.~Saelim, and P.~Tanedo, ``{The Same-Sign Dilepton
  Signature of RPV/MFV SUSY},''
  \href{http://dx.doi.org/10.1007/JHEP04(2013)077}{{\em JHEP} {\bfseries 1304}
  (2013) 077},
\href{http://arxiv.org/abs/1302.2146}{{\ttfamily arXiv:1302.2146 [hep-ph]}}.

\bibitem{CMS-PAS-SUS-13-010}
{ CMS} collaboration, ``{Search for RPV SUSY in the four-lepton final state},''
  \href{https://cds.cern.ch/record/1550552}{CMS-PAS-SUS-13-010} (2013).

\bibitem{ATLAS-CONF-2013-036}
{ ATLAS} collaboration, ``{Search for supersymmetry in events with four or more
  leptons in 21~fb$^{-1}$ of pp collisions at $\sqrt s = 8$~TeV with the ATLAS
  detector},'' \href{https://cds.cern.ch/record/1532429}{ATLAS-CONF-2013-036}
  (2013).

\bibitem{Chatrchyan:2013xsw}
{ CMS Collaboration} collaboration, S.~Chatrchyan {\em et~al.}, ``{Search for
  top squarks in R-parity-violating supersymmetry using three or more leptons
  and b-tagged jets},''
\href{http://arxiv.org/abs/1306.6643}{{\ttfamily arXiv:1306.6643 [hep-ex]}}.

\bibitem{Chatrchyan:2012sv}
{ CMS Collaboration} collaboration, S.~Chatrchyan {\em et~al.}, ``{Search for
  pair production of third-generation leptoquarks and top squarks in $pp$
  collisions at $\sqrt{s}=7$ TeV},''
  \href{http://dx.doi.org/10.1103/PhysRevLett.110.081801}{{\em Phys.Rev.Lett.}
  {\bfseries 110} (2013) 081801},
\href{http://arxiv.org/abs/1210.5629}{{\ttfamily arXiv:1210.5629 [hep-ex]}}.

\bibitem{Chatrchyan:2011cj}
{ CMS} collaboration, ``{Search for three-jet resonances in $pp$ collisions at
  $\sqrt s = 7$~TeV},''
  \href{http://dx.doi.org/10.1103/PhysRevLett.107.101801}{{\em Phys.Rev.Lett.}
  {\bfseries 107} (2011) 101801},
  \href{http://arxiv.org/abs/1107.3084}{{\ttfamily arXiv:1107.3084 [hep-ex]}}.

\bibitem{ATLAS:2012dp}
{ ATLAS Collaboration} collaboration, G.~Aad {\em et~al.}, ``{Search for pair
  production of massive particles decaying into three quarks with the ATLAS
  detector in $\sqrt{s}=7$ TeV $pp$ collisions at the LHC},''
  \href{http://dx.doi.org/10.1007/JHEP12(2012)086}{{\em JHEP} {\bfseries 1212}
  (2012) 086},
\href{http://arxiv.org/abs/1210.4813}{{\ttfamily arXiv:1210.4813 [hep-ex]}}.

\bibitem{Chatrchyan:2013izb}
{ CMS Collaboration} collaboration, S.~Chatrchyan {\em et~al.}, ``{Search for
  pair-produced dijet resonances in four-jet final states in pp collisions at
  sqrt(s) = 7 TeV},''
  \href{http://dx.doi.org/10.1103/PhysRevLett.110.141802}{{\em Phys.Rev.Lett.}
  {\bfseries 110} (2013) 141802},
\href{http://arxiv.org/abs/1302.0531}{{\ttfamily arXiv:1302.0531 [hep-ex]}}.

\bibitem{Alves:2011wf}
{ LHC New Physics Working Group} collaboration, ``{Simplified Models for LHC
  New Physics Searches},''
\href{http://arxiv.org/abs/1105.2838}{{\ttfamily arXiv:1105.2838 [hep-ph]}}.

\bibitem{Papucci:2011wy}
M.~Papucci, J.~T. Ruderman, and A.~Weiler, ``{Natural SUSY Endures},''
\href{http://arxiv.org/abs/1110.6926}{{\ttfamily arXiv:1110.6926 [hep-ph]}}.

\bibitem{Nikolidakis:2007fc}
E.~Nikolidakis and C.~Smith, ``{Minimal Flavor Violation, Seesaw, and
  $R$-parity},'' \href{http://dx.doi.org/10.1103/PhysRevD.77.015021}{{\em
  Phys.Rev.} {\bfseries D77} (2008) 015021},
\href{http://arxiv.org/abs/0710.3129}{{\ttfamily arXiv:0710.3129 [hep-ph]}}.

\bibitem{Csaki:2011ge}
C.~Csaki, Y.~Grossman, and B.~Heidenreich, ``{MFV SUSY: A Natural Theory for
  $R$-Parity Violation},''
  \href{http://dx.doi.org/10.1103/PhysRevD.85.095009}{{\em Phys.Rev.}
  {\bfseries D85} (2012) 095009},
\href{http://arxiv.org/abs/1111.1239}{{\ttfamily arXiv:1111.1239 [hep-ph]}}.

\bibitem{Krnjaic:2012aj}
G.~Krnjaic and D.~Stolarski, ``{Gauging the Way to MFV},''
  \href{http://dx.doi.org/10.1007/JHEP04(2013)064}{{\em JHEP} {\bfseries
  JHEP04} (2013) 064},
\href{http://arxiv.org/abs/1212.4860}{{\ttfamily arXiv:1212.4860 [hep-ph]}}.

\bibitem{Franceschini:2013ne}
R.~Franceschini and R.~Mohapatra, ``{New Patterns of Natural R-Parity Violation
  with Supersymmetric Gauged Flavor},''
  \href{http://dx.doi.org/10.1007/JHEP04(2013)098}{{\em JHEP} {\bfseries 1304}
  (2013) 098},
\href{http://arxiv.org/abs/1301.3637}{{\ttfamily arXiv:1301.3637 [hep-ph]}}.

\bibitem{Csaki:2013we}
C.~Csaki and B.~Heidenreich, ``{A Complete Model for R-parity Violation},''
\href{http://arxiv.org/abs/1302.0004}{{\ttfamily arXiv:1302.0004 [hep-ph]}}.

\bibitem{Barbier:2004ez}
R.~Barbier {\em et~al.}, ``{$R$-parity violating supersymmetry},''
  \href{http://dx.doi.org/10.1016/j.physrep.2005.08.006}{{\em Phys.Rept.}
  {\bfseries 420} (2005) 1},
\href{http://arxiv.org/abs/hep-ph/0406039}{{\ttfamily arXiv:hep-ph/0406039
  [hep-ph]}}.

\bibitem{Kao:2009fg}
Y.~Kao and T.~Takeuchi, ``{Single-Coupling Bounds on $R$-parity violating
  Supersymmetry, an update},''
\href{http://arxiv.org/abs/0910.4980}{{\ttfamily arXiv:0910.4980 [hep-ph]}}.

\bibitem{PY8}
T.~{Sj\"{o}strand}, S.~Mrenna, and P.~Skands, ``{A brief introduction to
  \textsc{Pythia}~8.1},''
  \href{http://dx.doi.org/10.1016/j.cpc.2008.01.036}{{\em Comput.Phys.Commun.}
  {\bfseries 178} (2008) 852}, \href{http://arxiv.org/abs/0710.3820}{{\ttfamily
  arXiv:0710.3820 [hep-ph]}}.
See~also~\url{http://home.thep.lu.se/~torbjorn/Pythia.html}.

\bibitem{deFavereau:2013fsa}
{ DELPHES 3} collaboration, J.~de~Favereau {\em et~al.}, ``{DELPHES 3, A
  modular framework for fast simulation of a generic collider experiment},''
  \href{http://dx.doi.org/10.1007/JHEP02(2014)057}{{\em JHEP} {\bfseries 1402}
  (2014) 057},
\href{http://arxiv.org/abs/1307.6346}{{\ttfamily arXiv:1307.6346 [hep-ex]}}.

\bibitem{Avetisyan:2013dta}
A.~Avetisyan, S.~Bhattacharya, M.~Narain, S.~Padhi, J.~Hirschauer, {\em
  et~al.}, ``{Snowmass Energy Frontier Simulations using the Open Science Grid
  (A Snowmass 2013 whitepaper)},''
\href{http://arxiv.org/abs/1308.0843}{{\ttfamily arXiv:1308.0843 [hep-ex]}}.

\bibitem{Avetisyan:2013onh}
A.~Avetisyan, J.~M. Campbell, T.~Cohen, N.~Dhingra, J.~Hirschauer, {\em
  et~al.}, ``{Methods and Results for Standard Model Event Generation at
  $\sqrt{s}$ = 14 TeV, 33 TeV and 100 TeV Proton Colliders (A Snowmass
  Whitepaper)},''
\href{http://arxiv.org/abs/1308.1636}{{\ttfamily arXiv:1308.1636 [hep-ex]}}.

\end{thebibliography}\endgroup

\end{document}